\documentclass[aps,prb,twocolumn,epsf]{revtex4}
\usepackage{graphicx}
\usepackage{graphics}
\usepackage{dcolumn}
\usepackage{amsmath}
\usepackage{amssymb}
\usepackage{bbold}
\usepackage{latexsym}
\usepackage{times}
\newcommand{\dg}{\dagger}
\newcommand{\up}{\uparrow}
\newcommand{\dwn}{\downarrow}

\newcommand{\N}{\hat N}

\newcommand{\ha}{\hat a}
\newcommand{\hb}{\hat b}
\newcommand{\hd}{\hat d}
\newcommand{\la}{\langle}
\newcommand{\ra}{\rangle}

\newcommand{\hth}{\hat \theta}

\begin{document}
\bibliographystyle{apsrev}
\title{Anomalous spin transport in a two-channel-Kondo quantum dot device}
\author{Prashant Sharma}
\affiliation{Laboratory of Atomic and Solid State Physics, Cornell
University, Ithaca NY 14853}
\date{\today}
\begin{abstract}
We study the response of a two-channel Kondo quantum dot device
proposed by Y. Oreg and D. Goldhaber-Gordon [Phys. Rev. Lett. {\bf
90}, 136602 (2003)] to a spin-bias applied across one of its channels
formed by Fermi liquid reservoirs weakly coupled to a spin-1/2 quantum
dot. When the temperature $T<T_K$, the Kondo temperature of the
device, the spin conductance depends on the Kondo coupling of the dot
spin to the other channel in an anomalous manner. For isotropic Kondo
couplings to the two channels the spin conductance is quantized for
$T\ll T_K$ characterizing the two-channel Kondo fixed point. On the
other hand, for anisotropic couplings a crossover energy scale
$T_A\neq 0$ determines the temperature $T\ll T_A$ when the spin
conductance vanishes indicating one-channel Kondo behavior.
\end{abstract}
\maketitle
\begin{figure}[t]
\includegraphics[bb=10 320 600 690,scale=.35]{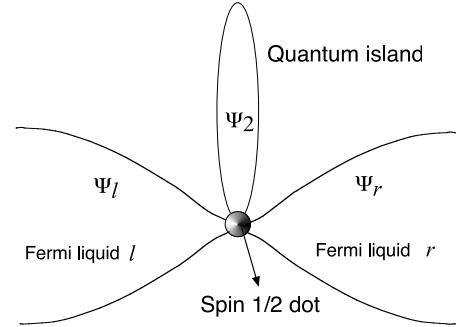}
\caption{Single electron transistor device of
Ref.~[\onlinecite{gordon03}]. The two channels that hybridize with the
dot are: (i) quantum island, and (ii) a linear combination of the
electron channels $\Psi_{l}$ and $\Psi_r$ of the two Fermi liquid
reservoirs.}\label{set-2ck}
\end{figure}

The Kondo effect~\cite{hewson,nozieres80} is a well understood problem
in condensed matter physics, and its experimental
realization~\cite{Goldhaber98, Cronenwett98,
Goldhaber98a,Wiel00,Nygard00,Liang02} in quantum dot systems allows
for a detailed investigation of several of its interesting theoretical
features~\cite{kaminski99,glazman01,affleck01} using transport
measurements. The low-bias transport properties of an odd-electron
quantum dot weakly coupled to two Fermi liquid reservoirs have been
well described by the physics near one-channel Kondo fixed
point~\cite{kaminski00}: When charge fluctuations in the dot can be
neglected, the odd-electron spin of the quantum dot hybridizes with a
single electron channel formed by a linear combination of electrons
from the two reservoirs. The physics of multi-channel Kondo has,
however, not been observed in such systems, although there have been a
few theoretical proposals for observing two-channel Kondo physics in a
non-equilibrium situation ~\cite{wen98,coleman01,rosch01}: When a
large source-drain bias introduces decoherence between electrons in
different reservoirs, so that they can be treated as two independent
channels interacting with the spin of the quantum dot. Perhaps the
most promising realization of the two-channel Kondo behavior in
equilibrium is provided by an experimental set-up in which the quantum
dot is coupled to two electron channels which have sufficiently strong
repulsive electron-electron interactions to inhibit transfer of
electrons from one channel to another.  The importance of
electron-electron interactions in stabilizing the two-channel Kondo
fixed point, in the case when the channels are two identical Luttinger
liquids, was first discussed in Ref.~\onlinecite{fabrizio95}, and
later, in the context of a quantum dot embedded in a carbon nanotube,
in Ref.~\onlinecite{kim02}. The recent experimental proposal by Oreg
and Goldhaber-Gordon, Ref.~\onlinecite{gordon03}, of a modified
single-electron transistor to observe two-channel Kondo physics, is
based on similar ideas in an experimentally realizable geometry: An
odd-electron quantum dot connected to a Fermi liquid reservoir, and
also to a larger quantum island (see Fig.~\ref{set-2ck}). A
sufficiently large charging energy $E_c$ of the quantum island
inhibits particle transfer to and from the Fermi liquid reservoir
leading to two-channel Kondo behavior~\cite{gordon03,florens03} when
$T\ll E_c$. This set-up allows for charge transport
measurements~\cite{pustilnik03}to be made using the Fermi liquid
reservoirs of the device.
Furthermore, as recent experiments reported in
Ref.~\onlinecite{marcus03} have realized a "spin battery", the
two-channel Kondo device can be probed by spin transport as well. The
"spin battery", which works by using adiabatic pumping of a chaotic
cavity, provides a source for spin-bias which is controlled by the
pumping frequency~\cite{mucciolo02} and by spin-orbit interaction
effects~\cite{sharma03} in the cavity.

The two-channel Kondo model that describes the low energy physics in
the above mentioned system has several distinctive {\it non}-Fermi
liquid features: (i) a logarithmic temperature dependence of the
specific heat and magnetic susceptibity, (ii) a non-zero ground state
entropy of magnitude $\frac{1}{2}\ln 2$.  A simple description that
captures these features is given in the language of abelian
bosonization~\cite{emery92,maldacena97,ye97,zarand00} by introducing
new spinor excitations that are non-locally related to the conduction
electrons. In the treatment of Ref.~\onlinecite{emery92}, the
Hamiltonian at the fixed point is written in terms of $C$-spinor
degrees of freedom and a local fermion $\hd$ representing the spin
impurity. The entropy of the ground state that develops below the
Kondo temperature, is associated with a local real (Majorana) fermion
$\hd+\hd^\dg$ that decouples from the conduction sea at the
two-channel Kondo fixed point. The $C$-spinor of the ``spin-flavour"
sector couples to the local Majorana fermion $i[\hd^\dg-\hd]$ only
through the linear combination $C_X+C_X^\dg$. The strength of this
coupling is determined by the Kondo temperature $T_K$. This effective
coupling implies that the phase variable $\hth_X$ (conjugate to the
difference in the number of spins between the two channels $\N_X$) is
fixed at the bottom of a cosine potential well with a depth $T_K$. As
a result, when the coupling is sufficiently strong ($T_K\gg T$),
eigenstates of the fixed point Hamiltonian contain phase coherent
superpositions of states with the same number of total spin but
different numbers of spins in the two channels~\cite{zarand00}.

In this paper anomalous effects are shown in spin transport in the
aforementioned quantum dot system, because of {\it non}-Fermi liquid
physics near the two-channel Kondo fixed point. The predicted effects
are: (i) A perfect spin conductance for isotropic Kondo coupling to
the two channels, when the charge conductance is made negligible by
asymmetrically coupling the right and left Fermi liquid reservoirs to
the quantum dot. (ii) A sharp change from spin conducting to spin
insulating behavior when the Kondo couplings are made anisotropic,
{\it i.e.}, there exists an anisotropy-dependent energy scale $T_A$
such that for $T\ll T_A$ the system has vanishing spin
conductance. Based on the picture mentioned above, a heuristic
understanding of this behavior may be arrived at as follows. When
$T_A=0$ and $\hth_X$ is a good quantum operator, a spin-bias $\lambda$
across one of the channels transforms the phase
$\hth_X\to\hth_X+\lambda t/2$. In the phase representation the current
operator is $\partial_t\hth_X/2\pi$, and we immediately obtain a
quantized spin-conductance in units of $\hbar/4\pi$.  In the presence
of anisotropy in the Kondo couplings, the ground state of the system
is qualitatively different: the strongly coupled channel forms a
singlet with the dot-spin, and $\hth_X$ is no longer a good quantum
operator. Therefore, the system is a spin insulator for energy scales
below $T_A$, determined by the channel anisotropy. In what follows, we
derive these results.

The modified single electron transistor can be described in terms of
electron channels in the two Fermi liquid reservoirs $\Psi_{l,r}(x)$
and in the large quantum island $\Psi_2(x)$ (see Fig.~\ref{set-2ck}),
that hybridize with the spin-1/2 quantum dot. As we are interested in
the low-energy physics, we linearize the electron excitation spectrum,
and using open boundary conditions at the site of the single electron
transistor ($x=0$) write the fields as chiral left-moving
fermions. Only a linear combination $\psi_1(x)=\Psi_{l}(x)\sin\theta_0
+ \Psi_r(x)\cos\theta_0$ couples to the spin of the quantum dot at
$x=0$, while the independent field $\psi_0(x)= \Psi_{l}(x)\cos\theta_0
- \Psi_r(x)\sin\theta_0$ decouples~\cite{pustilnik03}. The angle
$\theta_0=\tan^{-1}[t_l/t_r]$ is dependent on the (real) tunneling
amplitudes to the left ($t_l$) and right ($t_r$) Fermi liquid
reservoirs.

The low-energy behavior of the above mentioned device, when the charge
fluctuations in the quantum island are neglected, is that of the
two-channel Kondo problem~\cite{gordon03,florens03}. In terms of {\it
chiral} (left moving) Dirac fermions
$\psi_{1,2}(x)$~\cite{fabrizio95,zarand00} we can write the
Hamiltonian ($\hbar=1$):
\begin{eqnarray}\label{h2ck}
\cal{H}&=&\sum_{j=1,2}\sum_{\sigma=\up,\dwn}\int
dx\;\psi^\dg_{j\sigma}(x)\;iv\;\partial_x\psi_{j\sigma}(x) \nonumber\\
&-& \sum_{j=1,2}\sum_{\sigma\sigma'}J_{j}\vec
 S\cdot
\psi^\dg_{j\sigma'}(0)\frac{\vec\tau_{\sigma'\sigma}}{2}\psi_{j\sigma}(0).
\end{eqnarray}
Here, $\vec S$ represents the spin-1/2 degree of freedom of the
quantum dot, $\vec\tau$ are the Pauli matrices, and $v$ is the Fermi
velocity.
Applying a spin chemical potential difference $\lambda$ between the
right and left Fermi liquid leads adds a term
\begin{eqnarray}
H_\lambda&=&\lambda\sum_{\sigma=\up,\dwn}
\frac{\tau^3_{\sigma\sigma}}{2}
\left[N_{l\sigma}-N_{r\sigma}\right],\\
N_{l\sigma}&=&\int_{-L/2}^{L/2}dx\;\Psi_{l\sigma}(x)\Psi_{l\sigma}(x),
\end{eqnarray}
to the Hamiltonian ${\cal H}$ in Eq.~(\ref{h2ck}).  This can be
written in terms of the operator $\psi_1$ that couples to the spin of
the dot, and the free field $\psi_0$, using the relation:
\begin{eqnarray}
N_{l\sigma}-N_{r\sigma}&=&\cos2\theta_0\; [N_{1\sigma}-N_{0\sigma}] 
-\sin2\theta_0 N_{01\sigma},\\
N_{01\sigma}&=&\int dx[\psi_{0\sigma}^\dg(x)\psi_{1\sigma}(x)+H.c.].
\end{eqnarray}
Having chosen the spin quantization axis, we use abelian bosonization
to calculate the spin current. The four chiral fermions
$\psi_{j\sigma}(x)$ can be written in terms of chiral bosons
$\phi_{j\sigma}$:
\begin{eqnarray}
 \psi_{j\sigma}(x)&=&\frac{\chi_{j\sigma}}{\sqrt{2\pi a_0}}
 e^{-i\hth_{j\sigma}}e^{-i \N_{j\sigma}\frac{2\pi}{L}x}
 e^{-i\phi_{j\sigma}(x)}.
\end{eqnarray}
Here $v/a_0$ is the bandwidth, $\N_{j\sigma}$ counts the change in the
number of electrons with spin $\sigma$ in channel $j$ with respect to
a free electron ground state, $\hth_{j\sigma}$ is its conjugate phase
operator, $L/2$ is the length of each channel, and the cocycles
$\chi_{j\sigma}$ are required to satisfy the relations:
$\{\chi_{j\sigma},\chi_{j'\sigma'}\}=2\delta_{jj'}\delta_{\sigma\sigma'}$.
It is clear from the form of ${\cal H}$ that it can be written
entirely in the spin sector (involving only the bosonic fields
$\phi_{j\up}-\phi_{j\dwn}$). The only term that couples spin fields to
the charge fields ($\phi_{j\up}+\phi_{j\dwn}$) is the $N_{01\sigma}$
term in the external perturbation $H_\lambda$.

In the experimental set-up the parameter $\theta_0$ can be made
small. As a result it is convenient to write the spin current operator
$I_{\rm sp}$ as a sum of two distinct contributions:
\begin{eqnarray}\label{ispin-1}
I_{\rm sp} &=&\frac{1}{2}
\frac{d}{dt}\tau^3_{\sigma\sigma}[N_{l\sigma}-N_{r\sigma}]
=  I_{11} +  I_{01}\\
I_{11}
&=&\cos2\theta_0\frac{1}{2}\frac{d}{dt}\sum_\sigma\tau^3_{\sigma\sigma}
N_{1\sigma}\label{ispin}\\
I_{01}&=&\sin2\theta_0\frac{1}{2}\frac{d}{dt}\sum_\sigma
\tau^3_{\sigma\sigma}[\psi^\dg_{0\sigma}\psi_{1\sigma}+H.c.]
\end{eqnarray}
It is important to note that the two-channel Kondo
Hamiltonian~(\ref{h2ck}) has spin- and charge- sectors separated, and
also that the field $\psi_0$ does not couple to the dot
spin. Therefore, the spin current $\la I_{\rm sp}\ra$ can only depend
on even powers of $\sin2\theta_0$. This is easy to see for $\la
I_{11}\ra$, since $I_{11}$ depends only on the spin sector the lowest
order contribution is quadratic in $V\sin2\theta_0$. For $\la
I_{01}\ra$, the lowest order contribution is from $N_{01}$ term of
$H_\lambda$. This is apparently of order $V\sin^22\theta_0$, however,
it can easily be shown to vanish~\cite{pustilnik03}. It follows then,
that there is no contribution to the zero bias conductance upto terms
of order $\sin^22\theta_0$. Therefore, if one arranges the
experimental set-up such that $\sin\theta_0\ll 1$, we can neglect
these contributions altogether. As we show below, the remaining
dominant contribution to the spin current shows scaling behavior with
temperature and externally applied spin bias $\lambda$. The $\theta_0$
independent spin current is in marked contrast to the charge current
(in response to a charge-bias) that has been shown to have a peak
value of $2e^2/h\sin^22\theta_0$~\cite{pustilnik03}.
In what follows, we shall calculate the spin current $\la I_{\rm
sp}\ra$ in response to the externally applied bias
$
H_\lambda = \sum_\sigma \lambda\frac{\tau^3_{\sigma\sigma}}{2} N_{1\sigma}.
$
We begin by introducing {\it charge} (${\rm c}$), {\it spin} (${\rm
s}$), {\it flavour} (${\rm f}$), and {\it spin-flavour} (${X}$)
bosons:
\begin{eqnarray}
 \Phi_{\rm
 c(s)}&=&\frac{1}{2}\sum_{j=1,2}[\phi_{j\up}\pm\phi_{j\dwn}],\,
 \Phi_{{\rm
 f}}=\frac{1}{2}\sum_{\beta}[\phi_{1\beta}-\phi_{2\beta}],\nonumber\\
 \Phi_{X}&=&\frac{1}{2}[\phi_{1\up}-\phi_{1\dwn}-\phi_{2\up}+\phi_{2\dwn}].
 \end{eqnarray}
Similar relations hold for defining the corresponding zero-mode
number $\N_{\rm c,s,f,X}$ and phase operators $\hth_{\rm
c,s,{f},X}$. The physics near the two-channel-Kondo fixed point is
conveniently described in terms of a new basis of
$C$-spinors~\cite{emery92,maldacena97}: 
 \begin{eqnarray}
 C_{\alpha}(x)=\frac{\hat\chi_{\alpha}}{\sqrt{2\pi a_0}}
 e^{-i\hth_{\alpha}}e^{-i\N_{\alpha} 2\pi x/L}e^{-i\Phi_{\alpha}(x)},
 \end{eqnarray}
where $\alpha={\rm c, s, f, X}$. Following
Ref.~[\onlinecite{emery92}], a unitary transform $U={\rm
exp}[iS^z\Phi_{\rm s}(0)]$ yields the Hamiltonian ${\cal H}'=U{\cal
H}U^\dg$ which has spin (${\rm s}$) and spin-flavour (${X}$) sectors
decoupled from the charge (${\rm c}$) and flavour (${\rm f}$)
sectors. The Hamiltonian ${\cal H}'$ can be written in terms of the
$C$-spinors and the local fermion $\hd=\chi_{\rm s}e^{i\hth_{\rm
s}}S^-$, where $\hth_{\rm s}$ is the phase conjugate to the total spin
number operator $\N_{\rm s}$,
\begin{eqnarray}\label{hprime}
{\cal H}'&=&U{\cal H}U^\dg\nonumber\\&=& i\int
dx\;v\left[C_X^\dg(x)\partial_x C_X(x) + C_{\rm s}(x)\partial_x C_{\rm
s}(x)\right]\nonumber\\ &+&\Big\{\frac{J_1}{\sqrt{2\pi a_0}}\hd\;
C_X^\dg(0) + \frac{J_2}{\sqrt{2\pi a_0}}\hd\; C_X(0) +
H.c.\Big\}\nonumber\\ &+&\frac{1}{2}\left[J_1+J_2 - 4\pi
v\right][\hd^\dg\hd-\frac{1}{2}]:C_{\rm s}^\dg(0) C_{\rm
s}(0):\nonumber\\
&+&\frac{1}{2}\left[J_1-J_2\right][\hd^\dg\hd-\frac{1}{2}]:C_{X}^\dg(0)
C_{X}(0):
\end{eqnarray}

In the absence of channel anisotropy $J_1=J_2$, the above two-channel
Kondo system flows to a fixed point with the critical coupling
$[J_1+J_2]/2=J^*=2\pi v$,~\cite{fabrizio95,ye97} which gives the Kondo
temperature $J^*/\sqrt{2\pi a}\equiv\sqrt{T_{K}}$, where $v/a$ is the
reduced bandwidth. We restrict our attention to the effective
Hamiltonian ${\cal H}_{\rm 2CK}$ that contains all the relevant (in
the renormalization group sense) operators near the two-channel Kondo
fixed point~\cite{affleck92}. This can be conveniently written by
introducing Majorana (real) fermions
\begin{eqnarray}
\ha&=&\frac{\hd-\hd^\dg}{i\sqrt{2}},\,\,\hb=\frac{\hd+\hd^\dg}{\sqrt{2}}
\nonumber\\
\eta_a&=&\frac{[C_X(x)+C_X^\dg(x)]}{\sqrt{2}},\,\,
\eta_b=\frac{[C_X(x)-C_X^\dg(x)]}{\sqrt{2}}.
\end{eqnarray} 
The effective Hamiltonian:
\begin{eqnarray}
{\cal H}_{\rm 2CK}&=&\frac{1}{2}\int\!
dx\;\sum_{j=a,b}\eta_j(x)\left[iv{\partial_x}\right]\eta_j(x)
\nonumber\\
 &-&i 2\sqrt{T_{K}}\;\ha\;\eta_a(0) - i
2\sqrt{T_A}\;\hb\eta_b(0),
\label{s2ck}
\end{eqnarray}
valid for energy scales below the Kondo temperature $T_K$, also
contains the crossover energy scale associated with channel anisotropy
$T_A$. This energy scale is related to the Kondo couplings in
Eq.~(\ref{h2ck}) as~\cite{pustilnik03} $T_A\sim [2\pi
v(J_1-J_2)/(J_1+J_2)^2]^2 T_K$. The external spin-bias in the basis of
$C$-spinors
\begin{eqnarray}
H_\lambda'&=&UH_\lambda U^\dg=\frac{\lambda}{2}N_X - \frac{\lambda}{2} S^z,
\end{eqnarray}
where $N_X=\int dx :C_X^\dg(x)C_X(x):$ and $S^z=-i\ha\hb$ in terms of
the local Majorana fermions. The corresponding spin current operator
is
\begin{eqnarray}\label{ispinp}
I'_{\rm sp}=U I_{\rm sp}U^\dg
=\frac{1}{2}\frac{d}{dt}\left[N_X +i\ha\hb\right]
\end{eqnarray}
A transformation $C_X\to C_Xe^{i\lambda(x-vt)/2}$ removes the first
term of $H_\lambda'$, and corresponds to a change in the chemical
potential of the $C_X$ fermions. We will calculate the energy
dependent phase shift of these fermions as they scatter off the local
Majorana fermions. As these are left-movers we choose the incoming
states to the right of the impurity ($x>0$). The equations of motion
of these fermions are:
\begin{subequations}
\begin{eqnarray}
\partial_t\eta_{a}(x)&=&v\partial_x\eta_{a}(x)+\delta(x)\sqrt{T_K}\ha\\
\partial_t\ha&=&-\frac{1}{2}\lambda\hb - 2\sqrt{T_K}\eta_a(0)\\
\partial_t\hb&=&+\frac{1}{2}\lambda\ha + 2\sqrt{T_A}\eta_b(0)
\end{eqnarray}
\end{subequations}
Integrating across the impurity gives us:
\begin{subequations}\label{eqmotions}
\begin{eqnarray}
2\sqrt{T_K}\ha&=&\eta_{a}(0^-)-\eta_{a}(0^+)\\
2\sqrt{T_A}\hb&=&\eta_{b}(0^+)-\eta_{b}(0^-)\\
2\partial_t\ha&=&-\lambda\hb - 2\sqrt{T_K}[\eta_a(0^+)+\eta_a(0^-)]\\
2\partial_t\hb&=&\lambda\ha +
2\sqrt{T_A}[\eta_b(0^+)+\eta_b(0^-)]
\end{eqnarray}
\end{subequations}
Introducing the scattering matrix ${\cal S}$ allows us to write the
outgoing fields near the impurity
\begin{equation}
\eta_{Y=a,b}(0^-)=\sum_{Y'=a,b}\sum_k {\cal S}_{YY'}(k)\eta_{Y',k}
\end{equation}
where $\eta_{Y,k}$ is the $k$-th mode of the incoming field
$\eta_Y(0^+)$.  Using this {\it ansatz} in Eqs.~(\ref{eqmotions}) we
obtain:
\begin{eqnarray}
{\cal S}_{aa}&=&1+\frac{(e^{i\delta_a}-1)}
{\left[1 + \left(\frac{\lambda}{8\sqrt{T_K}}\right)^2
(e^{i\delta_a}-1)(e^{i\delta_b}-1)/T_A\right]}\\
{\cal S}_{ba}&=&\frac{\lambda}{2}\sqrt{\frac{T_A}{T_K}}
\frac{(e^{i\delta_b}-1)}{4T_A}{\cal S}_{aa},\\
e^{i\delta_a}&=&\frac{iv k+2T_K}{iv k - 2T_K},\,\,\,e^{i\delta_b}
=\frac{iv k+2T_A}{iv k - 2T_A}.
\end{eqnarray}
The scattering phase shift of the $C_X$ fermion at the two-channel
Kondo fixed point (when $T_A=0=\lambda$) is easily verified to be
$\delta_a(k=0)=\pi,\,\delta_b=0$, in accordance with
Ref.~[\onlinecite{maldacena97}]. The spin current~(\ref{ispinp}) can
be written in terms of the incoming and outgoing fields at the
impurity, using~(\ref{eqmotions}), and evaluated at finite bias and
temperature:
\begin{eqnarray}
\la I_{\rm sp}\ra&=&v\left\la C_X^\dg(0^+)C_X(0^+)
-C_X^\dg(0^-)C_X(0^-)\right\ra\nonumber\\
&=&\frac{v}{4\pi}\int dk\;{\rm Re}\left[{\cal S}_{aa}^*{\cal
S}_{bb}-1-|{\cal S}_{ab}|^2\right]\nonumber\\
&\times&\left[f(vk-\lambda/2)-f(vk+\lambda/2)\right].
\end{eqnarray}
Here $f(x)$ is the Fermi function, and the integral is limited by
the upper momentum cut-off $T_K/v$.
\begin{figure}[t]
\includegraphics[scale=.3]{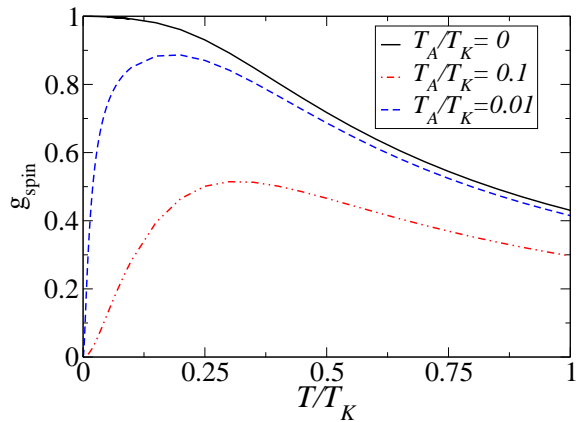}
\caption{Differential spin conductance $g_{\rm sp}$ as a function of
temperature for three different values of the channel anisotropy
energy scale $T_A$. Two-channel Kondo behavior is evident for $T_A\ll
T\ll T_K$, while one-channel Kondo behaviour for $T\ll T_A$.}
\label{gspin}
\end{figure}
At the two-channel Kondo fixed point ($T_A=0$) the current at low bias
 $\lambda\ll T_K$, 
 \begin{eqnarray}
\la I_{\rm sp}\ra&\equiv&\frac{\lambda}{4\pi}{g_{\rm sp}}
=-\frac{\lambda}{4\pi}\int dk\;\sin^2\left(\frac{\delta_a}{2}\right)
 \frac{df(vk)}{dk}\nonumber\\ 
{g_{\rm sp}}&=&\frac{2T_K}{\pi
T}\Psi'\left(\frac{1}{2}+\frac{2T_K}{\pi T}\right) 
\approx 1-\frac{1}{3}\left(\frac{\pi T}{2T_K}\right)^2.
\end{eqnarray}
Here $\Psi'(x)$ is the derivative of the psi function, and the last
expression is valid for $T\ll T_K$.  The quantized spin conductance
$g_{\rm sp}$ is the signature of the two-channel Kondo fixed
point. Deviations from the fixed point, because of $T_A\neq 0$ and
also because of an external magnetic field at the quantum dot site,
decrease the zero-bias spin conductance: At $T=0$ a non-zero $T_A$
leads to a vanishing zero-bias conductance, while
\begin{eqnarray}
g_{\rm sp}&=&-\int dk\;\sin^2\left(\frac{\delta_a-\delta_b}{2}\right)
 \frac{df(vk)}{dk}\nonumber\\
 &\approx&\left(1-\frac{T_A^2}{T_K^2}\right)
\zeta(2)\left(\frac{T}{T_A}\right)^2,
 \,{\rm for}\,T\ll T_A\ll T_K.
\end{eqnarray}
The full temperature dependence is plotted in Fig.~\ref{gspin}.

The role of local magnetic field is similar to that of channel
anisotropy. The zero-bias spin conductance in the presence of a local
magnetic field with Zeeman energy $\lambda_h$ is obtained by replacing
the spin-bias $\lambda\to\lambda_h$ in the phase shifts
$\delta_{a,b}$. The Zeeman energy corresponding to the crossover
energy $T_A$ is then found to be $\lambda_h\sim\sqrt{T_AT_K}$. Note
that we have shown the zero temperature behavior of spin conductance
to be identical to that of the impurity
entropy~\cite{gogolin95}. Therefore, the presence of a degenerate
ground state at the two-channel Kondo fixed point implies perfect
conductance. Also, note that the temperature dependence of the spin
conductance in this device, for $T_A=0$, is similar to that of charge
conductance in a one-channel Kondo set-up~\cite{kaminski00}, {\it
i.e.}, in the absence of the quantum island.
 
In conclusion, we have calculated the spin current through the
two-channel-Kondo quantum dot device of Ref.~\onlinecite{gordon03} for
temperature and spin-bias smaller than the Kondo temperature $T_K$. We
have shown that the spin conductance is simply related to the phase
shift of the $C_X$ spinor, and that the two-channel Kondo fixed point
can be identified by tuning the device parameters to obtain perfect
spin conductance. Deviations from the fixed point, because of channel
anisotropy and external magnetic field at the quantum dot site, are
shown to lead to a spin insulator.

This work was supported by the Packard foundation.


\end{document}